# Analyzing of MOS and Codec Selection for Voice over IP Technology

**Mohd Nazri Ismail**
**Department of MIIT,**
**University of Kuala Lumpur (UniKL), Malaysia**
mnazrii@miit.unikl.edu.my

**ABSTRACT**. In this research, we propose an architectural solution to implement the voice over IP (VoIP) service in campus environment network. Voice over IP (VoIP) technology has become a discussion issue for this time being. Today, the deployment of this technology on an organization truly can give a great financial benefit over traditional telephony. Therefore, this study is to analyze the VoIP Codec selection and investigate the Mean Opinion Score (MOS) performance areas evolved with the quality of service delivered by soft phone and IP phone. This study focuses on quality of voice prediction such as i) accuracy of MOS between automated system and human perception and ii) different types of codec performance measurement via human perception using MOS technique. In this study, network management system (NMS) is used to monitor and capture the performance of VoIP in campus environment. In addition, the most apparent of implementing soft phone and IP phone in campus environment is to define the best codec selection that can be used in operational environment. Based on the finding result, the MOS measurement through automated and manual system is able to predict and evaluate VoIP performance. In addition, based on manual MOS measurement, VoIP conversations over LAN contribute more reliability and availability performance compare to WAN.

## Introduction

As with most new technologies, Voice over Internet Protocol (VoIP) brings new challenges along with the benefits. The main challenge is VoIP's extreme sensitivity to delay and packet loss compared with other network applications such as web and e-mail services. A basic understanding of VoIP traffic and of the quality metrics provided by VoIP monitoring tools will help to keep VoIP network running smoothly. Existing hardware video phones have integrated video camera and microphone and can be used independently of any other equipment [VCC06]. Traditionally, set top boxes are video broadcast receivers and are equipped with video/audio decoder (MPEG2) and usually based on H.263 codec for video and G.711, G.723 or G.729 codecs for voice [VCC06]. The actual multimedia content (voice and video) is usually transmitted by means of the Real-time Transport Protocol (RTP) [S+03]. The objectives of this study as follows: i) design and implement VoIP technology for ease interactions and communications between UniKL branches; ii) provide toll free communication platform using real time voice communication in campus environment; iii) implement and test the accuracy of VoIP management system with manual system using MOS technique; and iv) use and test different types of codec for VoIP conversation.

This study focuses on quality of voice and video performance using MOS technique such as i) automated MOS measurement; ii) physical (manual) MOS by human perception measurement; and iii) different codec selection measurement. Today, many people are talked and communicated face to face using VoIP Phone system. When VoIP is implemented using the public Internet, users may experience quality degradations due to dynamic delays and losses in the LAN and Internet. Packets may be lost, either in isolation or in batches, and may experience sudden delay increases [SW07]. Figure 1 and Figure 2 show USB Phone system and soft phone application will use in real network environment for the experimental. In addition, Figure 3 shows the sample of VoIP network architecture.

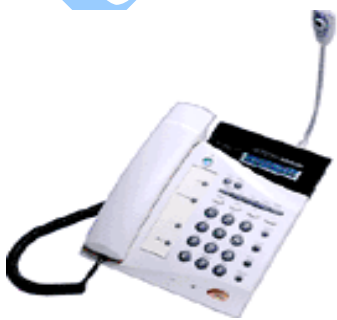 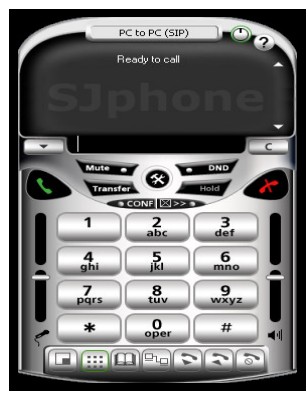

Fig. 1: USB Video Phone System    Fig. 2: Soft Phone Application

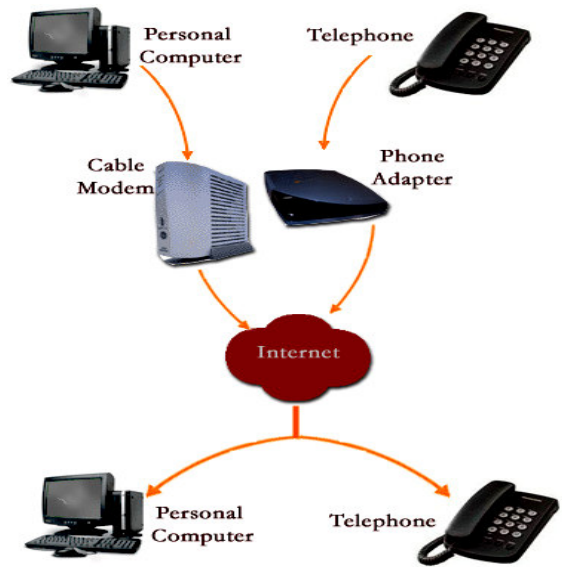

Fig. 3: Sample of VoIP Architecture

## 1. Related Works

Recently, VoIP (Voice over IP) [Col02] is rapidly growing and becoming a mainstream telecommunication service, not only because of the lower cost compared with traditional PSTN (Public Switched Telephone Network), but also its convergence technologies of data and voice communication. VoIP applications like Skype [BS06] have also achieved great success. However, due to the complexity of the Internet, it is unpractical to calculate VoIP performance metrics only through the mathematical modeling, as what was done in the telephone networks, so the performance evaluation of VoIP requires actual measurement activities.

There have been numerous studies on VoIP measurement. A. Markopoulou [MTK02] measured loss and delay characteristics of American backbone networks, and analyzed how these characteristics impact VoIP quality. For examples, most work focused on monitoring and analyzing performance of actual applications, like MSN and Skype [C+06a] [C+06b]. Skype using more to video codec like H.263 and H.261. Skype also using the audio codec likes G.711, G.729, G.723, G.728 and GSM. In contrast to other works in the literature, we implemented VoIP network environment to measure the performance of VoIP using MOS technique at campus environment.

## 2. Methodology

Figure 4 shows the overall framework of the VoIP service in campus environment. There are five phases development process such as: i) planning and research; ii) development; iii) implementation; iv) testing and v) documentations.

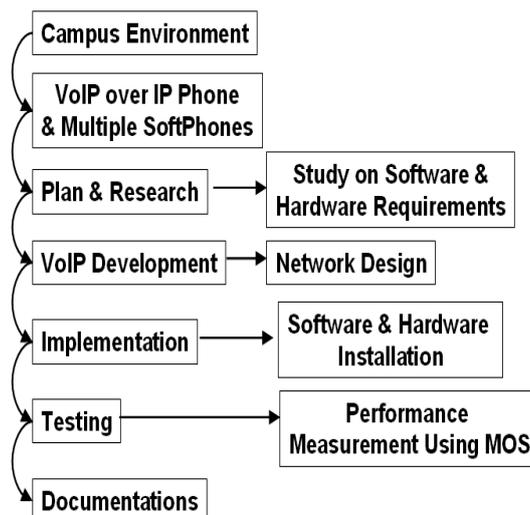

Fig. 4: Framework of VoIP Service in Campus Environment Development

Figure 5 shows the overall framework of the VoIP performance analysis using MOS technique. In the experiment, the performance analysis will focus on automated MOS system, manual MOS measurement and different codec selection over LAN and WAN. Network management system such as VQ manager is used to analyze VoIP service via automated MOS measurement in campus environment. Manual MOS measurement is based on human perception.

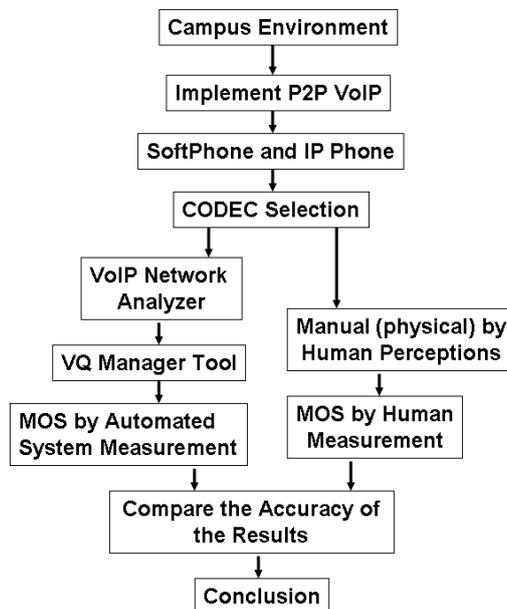

Fig. 5: Framework of VoIP Performance Analysis using MOS Technique

## 3. Proposed VoIP Network Architecture in Campus Environment

We have setup a real network environment to analyze and measure implementation of VoIP service at University of Kuala Lumpur (UniKL) in Malaysia. This study posits several research questions: i) what is the performance level of the VoIP over LAN and WAN using MOS technique; and ii) Is the analysis for evaluating and measuring VoIP performance effective using MOS technique?

Figure 6 shows the VoIP architecture in real network campus environment. VoIP quality can be monitored periodically through the measurement using VQnet (VoIP) management tools to gather quality variation information, avoiding the ignorance of unacceptable VoIP quality caused by the network failure or bandwidth bottleneck (see Figure 7). Figure 8 shows MOS technique implementation in campus environment using IP phone and multiple soft phones performance measurement. Several soft phones have used to evaluate and measurement of VoIP performance using MOS technique such as i) Wengo Phone; ii) 3CX; iii) PhoneLite; iv) Kapanga; v) SJPhone and others [BHW07].

267

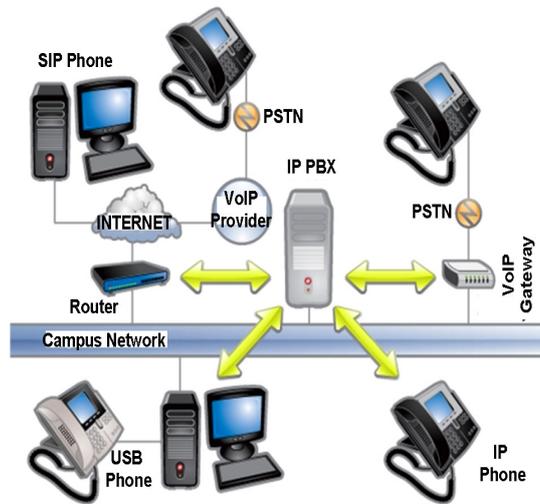

Fig. 6: Development of V2oIP Architecture in Real Network

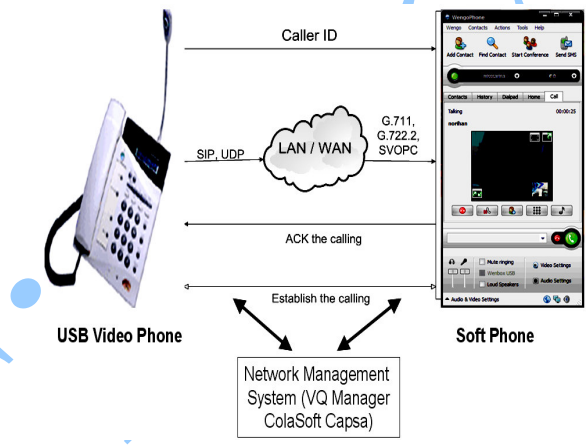

Fig. 7: Installation of Network Monitoring System for VoIP

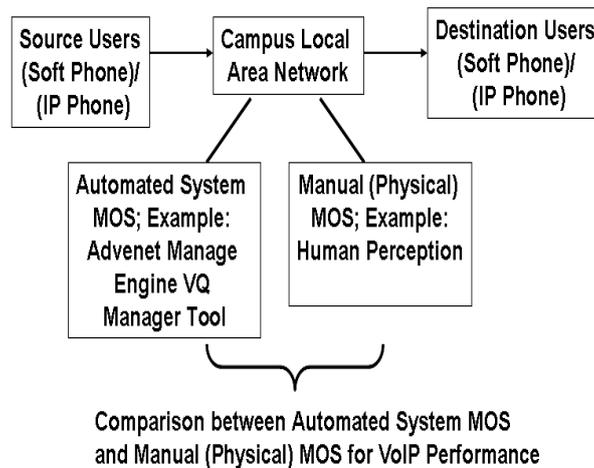

Fig. 8: Implementation of MOS Technique

## 4. Experimental and Analysis Results

This section analyses, measures and compares VoIP performance in campus environment. The MOS is expressed as a single number in the range 1 to 5, where 1 is lowest perceived quality, and 5 is the highest perceived quality. The MOS is generated by averaging the results of a set of standard, subjective tests where a number of listeners rate the heard audio quality of test sentences read aloud by both male and female speakers over the communications medium being tested. In the VoIP world, voice quality is measured by a 'Mean Opinion Score', which is a number between 1 and 5 used to quantitatively express the subjective quality of speech in communications systems, especially digital networks that carry VoIP traffic. Anything above a 4.0 is considered toll grade (see Figure 9 and Table 1).

Table 1: MOS Rating Measurement

| MOS | QUALITY | IMPAIRMENT |
|-----|---------|------------|
| 5 | Excellent | Imperceptible |
| 4 | Good | Perceptible but not annoying |
| 3 | Fair | Slightly annoying |
| 2 | Poor | Annoying |
| 1 | Bad | Very Annoying |

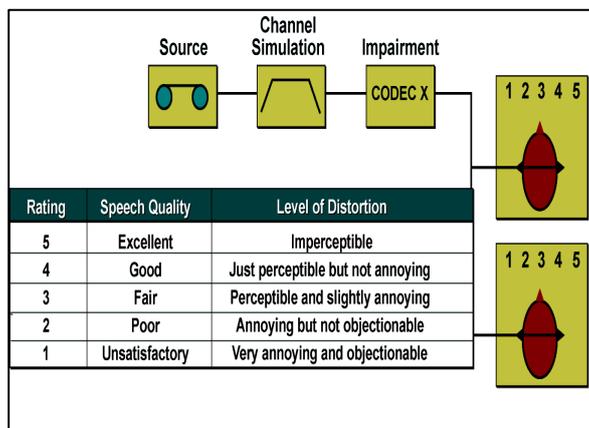

Fig. 9: MOS Technique Rating

*A. Comparison of MOS Automated System and Manual Human Perception*
Figure 10 shows the sample of MOS automated system that generate by 'Advenet Manage Engine VQ Manager' VoIP management tool. Soft and IP phones are configured using Codec G.711. We used several soft phones such as SJPhone and Wengo Phone to measure the accuracy of automated system and human perception.

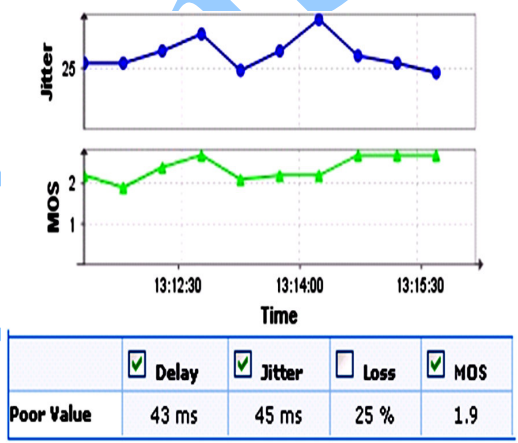

Figure 10: MOS Generate by Automated System

Based on the results, Figure 11 and Figure 12 show the MOS value between automated system and manual system (physical). Both automated and manual system measurement behave approximately the same graph pattern during the evaluation of VoIP performance. We conclude that base

270

on our findings, the MOS measurement through automated and manual system is able to predict and evaluate VoIP performance.

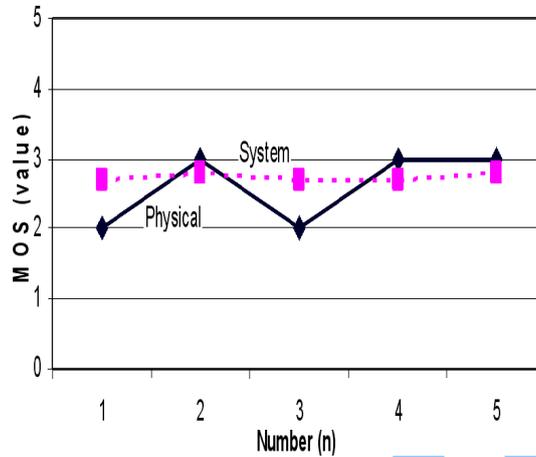

Figure 11: Comparison of MOS Automated and Manual System (Physical) via Wengo Phone Application

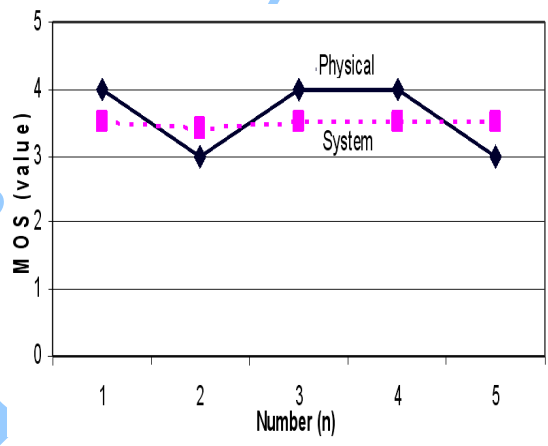

Figure 12: Comparison of MOS Automated and Manual System (Physical) via SJPhone Application

**Different Codec Selection Performance Measurement by MOS Technique**: Codecs generally provide a compression capability to save network bandwidth. Some Codecs also support silence suppression, where silence is not encoded or transmitted. . The most common voice CODECs includes G.711, G.723, G.726, G.728, and G.729. Table 2 shows VoIP Codec protocols and algorithms. G.711 has two versions called U-law (US,

271

Japan) and A-law (Europe). U-law is indigenous to the T1 standard used in North America and Japan. In our experiments, we used four different types of Codec such as i) G.711 U-law; ii) G.711 A-law; iii) ILBC; and iv) speex. We selected five users to measure VoIP performance through MOS technique. The users need to make conversation between them.

Table 2: VoIP Codec Protocols and Algorithms

| Codec | Algorithm | Kbps |
|---|---|---|
| G.711 | PCM (Pulse Code Modulation) | 64 |
| G.722 | SBADPCM (Sub-Band Adaptive Differential Pulse Code Modulation) | 48, 56 and 64 |
| G.723 | Multi-rate Coder | 5.3 and 6.4 |
| G.726 | ADPCM (Adaptive Differential Pulse Code Modulation) | 16, 24, 32, and 40 |
| G.727 | Variable-Rate ADPCM | 16-40 |
| G.728 | LD-CELP (Low-Delay Code Excited Linear Prediction) | 16 |
| G.729 | CS-ACELP (Conjugate Structure Algebraic-Code Excited Linear Prediction) | 8 |
| ILBC | Internet Low Bitrate Codec | 13.33 and 15.20 |
| Speex | CELP (Code Excited Linear Prediction) | 2.15-44.2 |
| GSM | RPE-LTP (Regular Pulse Excitation Long-Term Prediction) | 13 |

We conclude that base on our findings, the MOS measurement through LAN and WAN using different types of Codec during conversation VoIP (see Table 3). Codec G.711 contributes high reliability VoIP conversation compare to iLBC and Speex. Both medium (LAN and WAN) contribute different performance and characteristics. If the PC (install soft phone) load is very high, it can achieve higher CPU usage and make the throughput low that can cause inconsistent conversation between two parties. Therefore, it is important to understand how delay, jitter and CPU usage can affect speech quality through soft phone. In addition, poor link quality network between two parties can also increase the delay and jitter for VoIP service. The efficiency of VoIP service in campus environment can quickly deteriorates when the jitter and delay increases in WAN (see Figure 13). The results of this study show that voice service is able to contribute and achieve reliability and scalability in LAN (see Figure 13). Therefore, proper plan can mitigate most of the negative effects that are related to V2oIP characteristics.

Table 3: Manual MOS Codec Measurement

| USERS | CODECS | MOS in LAN | MOS in WAN |
|---|---|---|---|
| User 1 | G711 aLaw | 4 | 3 |
| | G711 uLaw | 4 | 3 |
| | iLBC | 3 | 2 |
| | Speex | 3 | 2 |
| User 2 | G711 aLaw | 4 | 3 |
| | G711 uLaw | 3 | 2 |
| | iLBC | 2 | 1 |
| | Speex | 3 | 2 |
| User 3 | G711 aLaw | 3 | 2 |
| | G711 uLaw | 4 | 3 |
| | iLBC | 2 | 1 |
| | Speex | 3 | 2 |
| User 4 | G711 aLaw | 3 | 2 |
| | G711 uLaw | 4 | 3 |
| | iLBC | 3 | 2 |
| | Speex | 3 | 2 |
| User 5 | G711 aLaw | 4 | 3 |
| | G711 uLaw | 3 | 2 |
| | iLBC | 2 | 1 |
| | Speex | 2 | 1 |

**Conclusion**

Convergence of video, voice and data over IP with current network infrastructure can contribute many revenues to campus users. This paper discussed which
- MOS system (automated and manual measurement) could produce good performance evaluation;
- different types of VoIP Codec selection during conversation.

We conclude that base on our findings, WAN can contribute higher delay, packet loss and CPU usage compare to LAN in campus environment. While, automated management and manual system measurement will generate approximately similar results. If VoIP want to implement in campus environment, it is recommended to enable QoS function in order that to achieve a good quality conversation between two parties over LAN and WAN.

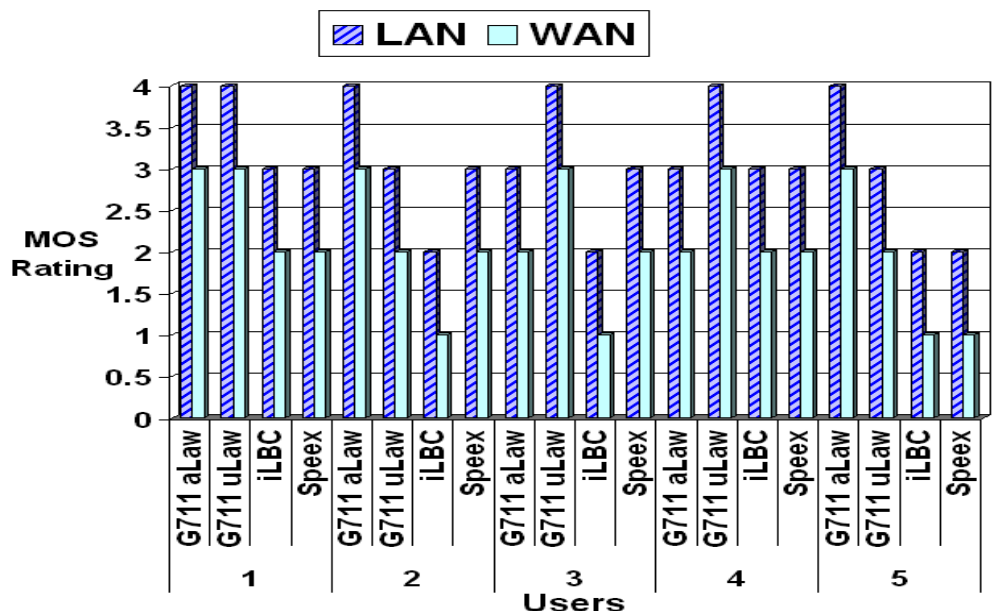

Figure 13: Comparison of VoIP over LAN and WAN Measurement

In future work, it should consider implementing techniques to improve quality of VoIP over WAN. Therefore, efficient and effective QoS provisioning techniques are important to achieve a robust service. There are several techniques should be studied and analyzed that can be used to increase performance of VoIP in campus environment as follows:
- Dejitter buffer;
- Type of Service (ToS);
- Weighted Fair Queuing (WFQ);
- Random Early Detection (RED).

Implementing quality of service mechanisms on peak hours is a method to improve V2oIP network service performance in campus environment.

**References**

[BS06]  S. A. Baset, H. Schulzrinne - *An analysis of the Skype peer-to-peer internet telephony protocol*, In Proc. of IEEE INFOCOM, Barcelona, Spain, Apr. 2006